\documentstyle[sprocl,epsf]{article}

\def\be{\begin{equation}}
\def\ee{\end{equation}}
\def\bea{\begin{eqnarray}}
\def\eea{\end{eqnarray}}
\begin{document}

\title{CENTER VORTEX MODEL FOR THE INFRARED SECTOR OF YANG-MILLS 
THEORY\footnote{Talk presented by M.~Engelhardt.} }

\author{M. ENGELHARDT,\footnote{email: 
\verb+engelm@pion08.tphys.physik.uni-tuebingen.de+ } H. REINHARDT}

\address{Institut f\"ur theoretische Physik, Universit\"at T\"ubingen,\\
Auf der Morgenstelle 14, 72076 T\"ubingen, Germany}

\author{M. FABER}

\address{Institut f\"ur Kernphysik, Technische Universit\"at Wien,\\
A-1040 Vienna, Austria}

\maketitle\abstracts{A model for the infrared sector of $SU(2)$ Yang-Mills 
theory, based on magnetic vortices represented by (closed) random surfaces, 
is presented. The random surfaces, governed by an action penalizing
curvature, are investigated using Monte Carlo methods on a hypercubic
lattice. A low-temperature confining phase and a high-temperature
deconfined phase are generated by this simple dynamics.
After fixing the parameters of the model such as to reproduce the
relation between deconfinement temperature and zero-temperature
string tension found in lattice Yang-Mills theory, a surprisingly
accurate prediction of the spatial string tension in the deconfined
phase results. Furthermore, the Pontryagin index associated with the 
lattice random surfaces of the model is constructed. This allows to also 
predict the topological susceptibility; the result is compatible with
measurements in lattice Yang-Mills theory. Thus, for the first time
an effective model description of the infrared sector emerges which 
simultaneously and consistently reproduces both confinement and the 
topological aspects of Yang-Mills theory within a unified framework.
Further details can be found in the e-prints hep-lat/9912003 and
hep-lat/0004013, both to appear in Nucl. Phys. B.}

\section{Motivation}
\label{intro}
Diverse nonperturbative phenomena characterize strong interaction physics.
Color charge is confined, chiral symmetry is spontaneously broken, and
the axial $U(1)$ part of the flavor symmetry exhibits an anomaly.
In principle, a theoretical tool exists which allows to calculate any
observable associated with these phenomena, namely lattice gauge theory.
Nevertheless, it is useful to concomitantly formulate effective models
which concentrate on the relevant infrared degrees of freedom, and thus
provide a clearer picture of the dominating physical mechanisms. This
facilitates the exploration of problem areas which are difficult to 
access using the full (lattice) gauge theory.

Indeed, such model descriptions of the infrared sector exist. The most
widely accepted picture of the confinement phenomenon at present is
the dual superconductor mechanism,\cite{mag} based on Abelian monopole
degrees of freedom. On the other hand, the $U_A (1)$ anomaly is most 
conveniently encoded in instanton degrees of freedom, which carry the 
relevant topological charge. Instanton models \cite{inst} are furthermore 
successful at modeling the spontaneous breaking of chiral symmetry. They, 
however, do not provide confinement, at least on the level of the simple 
model dynamics hitherto explored. Conversely, the connection of the 
aforementioned Abelian monopole degrees of freedom with the topological
and chiral properties of the underlying theory, QCD, has not been 
completely clarified. Thus, the different nonperturbative phenomena have
been explained on a separate footing, and no consistent, comprehensive
picture of the infrared regime has emerged.

The vortex model presented in the following aims to bridge this gap.
It simultaneously provides a quantitative description of both the
confinement properties of $SU(2)$ Yang-Mills theory (including the 
finite-temperature transition to a deconfined phase), as well as the
topological susceptibility, which encodes the $U_A (1)$ anomaly.
This turns out to be possible on the basis of a simple effective
model dynamics, detailed further below. At this stage, the chiral
condensate, which represents the order parameter for the spontaneous
breaking of chiral symmetry, has not been evaluated within the model.
Some remarks on this next important point of investigation follow in
section \ref{outsec}.

\section{Chromomagnetic center vortices and their dynamics}
\label{defsec}
Center vortices are closed lines of chromomagnetic flux in three-dimensional
space; thus, they are described by closed two-dimensional world-surfaces
in four-dimensional space-time. Their magnetic flux is quantized such that 
they contribute a phase corresponding to a nontrivial center element
of the gauge group to any Wilson loop they are linked to (or, equivalently,
whose minimal area they pierce). In the case of $SU(2)$ color discussed
here, the only such nontrivial center phase is $(-1)$. For higher gauge
groups, one must consider several possible fluxes carried by vortices.

As an illustrative example, consider a vortex surface located on the
$x_1 =x_2 =0$ plane, i.e. extending into the 3-4 direction. Such a
vortex surface can be associated with a gauge field as
follows.\footnote{It is possible to construct explicit gauge field
representations for arbitrary vortex surface configurations.\cite{cont} }
Describe the 1-2 plane in polar coordinates, 
$(x_1 ,x_2 ) \rightarrow (r,\varphi )$, and set the $\varphi $-component 
of the gauge field to
\begin{equation}
A_{\varphi } = \frac{\sigma_{3} }{2r}
\label{voradef}
\end{equation}
where $\sigma_{3} $ denotes the third Pauli matrix, encoding the color
structure. All other components of the gauge field can be set to zero.
Evaluating a circular Wilson loop of radius $R$ centered at $r=0$
yields
\begin{equation}
W [R,0] = \frac{1}{2} \mbox{Tr}
\exp \left( i\int_{0}^{2\pi } d\varphi \, R \ \frac{\sigma_{3} }{2R} \right)
= \frac{1}{2} \mbox{Tr} \exp \left( i\pi \sigma_{3} \right) = -1
\label{wcalc}
\end{equation}
in accordance with the defining property of a vortex given above. Note
that the result is independent of $R$; the flux of the vortex thus is
localized at $r=0$, and for any $R$, the full flux is encircled by
the Wilson loop. This vortex is infinitely thin. The real physical
vortex fluxes conjectured to describe the infrared aspects of the Yang-Mills
ensemble within the vortex picture of course should be thought of as
possessing a finite thickness;\footnote{Phenomenologically,
such a physical thickness has e.g.~been argued to be crucial for an
explanation of the Casimir scaling behavior of adjoint representation
Wilson loops.\cite{casscal} } however, for the purpose of evaluating 
deeply infrared observables, i.e. looking from far away, the thin 
idealization is adequate. Nevertheless, the thickness of the vortices
will significantly influence the ansatz for the vortex dynamics presented
further below, and the physical interpretation of that ansatz.

Having described the vortex by a gauge field, one can of course also
associate a field strength with it. In view of the Lorentz structure
of the field (\ref{voradef}), the only nonvanishing tensor component
of the field strength is $F_{r\varphi } $, i.e. the component
associated with the two space-time directions perpendicular to the
vortex. Lastly, it is important to note that an orientation can be
associated with the vortex surface, corresponding to the two possible
directions of the magnetic flux. Inverting the orientation simply means
changing the sign of the gauge field, $A \rightarrow -A$. If one 
inserts the oppositely oriented vortex field in (\ref{wcalc}), one
again obtains the value (-1) for the Wilson loop, i.e. the field $-A$
indeed also describes a vortex. Wilson loops, and thus the confinement
properties of a vortex ensemble, are insensitive to the orientations of
the vortices. However, since $A \rightarrow -A$ implies $F \rightarrow -F$,
topological properties will in general turn out to be sensitive to the
vortex orientations.

It should be noted that the chromomagnetic vortex fluxes described above
correspond precisely to the flux domains found in the so-called Copenhagen 
vacuum.\cite{spag} Thus, the idea that these degrees of freedom may be 
relevant in the infrared sector of Yang-Mills theory is not new. What is new 
about the model to be discussed here is that the vortex picture has been 
developed into a quantitative tool, allowing to evaluate diverse physical 
obervables, that the vortex picture furthermore has been generalized to 
finite temperatures, including the deconfined phase, and, perhaps most
importantly, that it has been extended to also subsume the topological
properties of the Yang-Mills ensemble.

There are two complementary ways to think about vortex dynamics, i.e. to
generate a vortex ensemble. On the one hand, one can devise an algorithm 
by which it is possible to identify and extract vortex structures from 
Yang-Mills field configurations.\cite{centg} Then, using the full 
Yang-Mills ensemble, as obtained e.g. in a standard lattice Monte Carlo 
experiment, in conjunction with the aforementioned algorithm, one can study 
the vortex physics induced by the full Yang-Mills dynamics. One drawback of 
such an approach is that no simplification is achieved compared with the
standard lattice gauge calculation; rather, it allows a physical
interpretation of the results of that calculation. This avenue will
not be pursued here. Instead, vortices will be the only degrees of
freedom entering the description from the very beginning, and a simple
effective model dynamics for the vortices will be assumed, with the
aim of ascertaining how far such a simple model will carry.

Specifically, to arrive at a tractable model, the vortex world-surfaces
will be composed of elementary squares (plaquettes) on a hypercubic
lattice. The spacing of this lattice will be a fixed physical quantity
related to the thickness of the vortex fluxes already mentioned further
above; the lattice prevents an arbitrarily close packing of the vortices. 
Thus, it is not envisaged to eventually take the lattice spacing to zero, 
and accordingly renormalize the coupling constants, such as to arrive at a 
continuum theory. Rather, the lattice spacing represents the fixed physical
cutoff one expects to be present in any infrared effective theory,
and thus also delineates the ultraviolet limit of validity of the model.
If one wants to refine the model such as to eliminate the artificial
hypercubic nature of the vortex surfaces, one has to replace the lattice
spacing by some other ultraviolet cutoff. For instance, if the surfaces
are represented as triangulations, a minimal area of the elementary
triangles could take on this role.

On the hypercubic lattice adopted here, the vortex surfaces will be
regarded as random surfaces. This is not such a far-fetched idea;
heuristically, the vortex degrees of freedom are magnetic degrees of
freedom, dual to the electric degrees of freedom in terms of which one
initially defines gauge theories.\footnote{This is particularly manifest
in the fact that the lattice on which center vortices are defined is
dual to the one on which standard lattice gauge theory is constructed.}
Thus, in the infrared regime, where the usual electric degrees of freedom
become strongly coupled, one would conversely expect magnetic vortex degrees
of freedom to be weakly coupled, i.e. weakly correlated. The reason the
magnetic degrees of freedom are nevertheless organized into vortex lines
lies in the constraint of continuity for the magnetic flux (which represents
the dual to the usual Gau\ss \ law constraint).

The random surfaces describing the vortices will be generated by Monte
Carlo methods; the weight function specifying the ensemble depends on
the curvature of the surfaces as follows. Every instance of a link on the
lattice being common to two plaquettes which are part of a vortex surface,
but which do not lie in the same plane, is penalized by an action 
increment $c$. Thus the action can be represented pictorially as
\vspace{-0.25cm}

\begin{figure}[h]
\hspace{6.5cm}
\epsfxsize=1cm
\epsffile{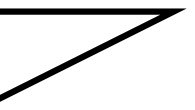}
\end{figure}
\vspace{-1.58cm}
\begin{equation}
S_{curv} = c \cdot \mbox{ {\LARGE \# ( \ \ \ ) } }
\label{actio}
\end{equation}
Note that several such pairs of plaquettes can occur for any given
link. For instance, if all six plaquettes attached to a given link
are part of a vortex surface, this implies a contribution $12c$
to the action.

Since this construction seems rather ad hoc, a comment is in order at
this point. As the random surface model discussed here is meant to
represent an infrared effective theory, it seems natural to expand the
action in a gradient expansion. In such an expansion, one would expect
a Nambu-Goto term as the leading term, i.e. a contribution to the action 
proportional to the total surface area. Only in the next to leading 
order, curvature terms appear. Such a generalized model has indeed
been investigated,\cite{selprep} containing both a Nambu-Goto action
as well as the curvature term $S_{curv} $ given above. It turns out
that, to a large extent, one can trade the effects of the two action
terms off against one another; more precisely, in the plane of
coupling coefficients corresponding to the two terms, one finds
lines on which confinement physics (specifically all string tensions,
as functions of temperature), as discussed in the next section, is 
invariant up to ten percent effects. This is not entirely surprising, 
since surfaces with a large area typically will also contain a large 
amount of curvature, and vice versa (up to exceptions which are 
entropically suppressed). Thus, both coupling coefficients to some extent 
have similar effects. The aforementioned lines of approximately invariant 
physics happen to cross the axis on which the Nambu-Goto coefficient 
vanishes (but not the axis on which the curvature coefficient $c$ does). 
Thus, one can restrict oneself to the models given by the action 
(\ref{actio}) without loss of generality.

As a last remark, for the purpose of studying topological properties,
cf. section \ref{topsec}, it is necessary to also specify orientations
for the vortex surfaces (this was already hinted at above). 
Section \ref{topsec} contains more details on this point.

\section{Confinement and Deconfinement}
Given the random surface dynamics defined in the previous section, it is
now straightforward to evaluate Wilson loops, using the fundamental
property of vortices that they modify any Wilson loop by a phase factor
$(-1)$ whenever they pierce its minimal area. Such measurements can
furthermore be carried out at several temperatures, by adjusting the
extension of (Euclidean) space-time in the time direction; this extension
is identified with the inverse temperature of the ensemble. At finite
temperatures, static quark potentials are given by Polyakov loop correlators;
their properties in the presence of vortex fluxes are completely 
analogous to the properties of Wilson loops. Qualitatively,
as long as the curvature coefficient $c$ is not too large, one finds a 
confined phase (non-zero string tension) at low temperatures, and a phase 
transition to a high-temperature deconfined phase. In order to make the
correspondence to full $SU(2)$ Yang-Mills theory quantitative, one can
adjust $c$ such as to reproduce the ratio of the deconfinement
temperature to the square root of the zero-temperature string tension,
$T_C /\sqrt{\sigma_{0} } =0.69$. This happens at the value $c=0.24$.
Furthermore, by setting $\sigma_{0} =(440 \, \mbox{MeV} )^2 $ to fix
the scale, one extracts from the measurement of $\sigma_{0} a^2 $ the
lattice spacing $a=0.39 \, \mbox{fm} $. As discussed in the previous
section, this is a fixed physical quantity, related to the thickness
of the vortices, which represents the ultraviolet limit of validity
of the effective vortex model.

\begin{figure}[t]
\centerline{
\epsfxsize=7cm
\epsffile{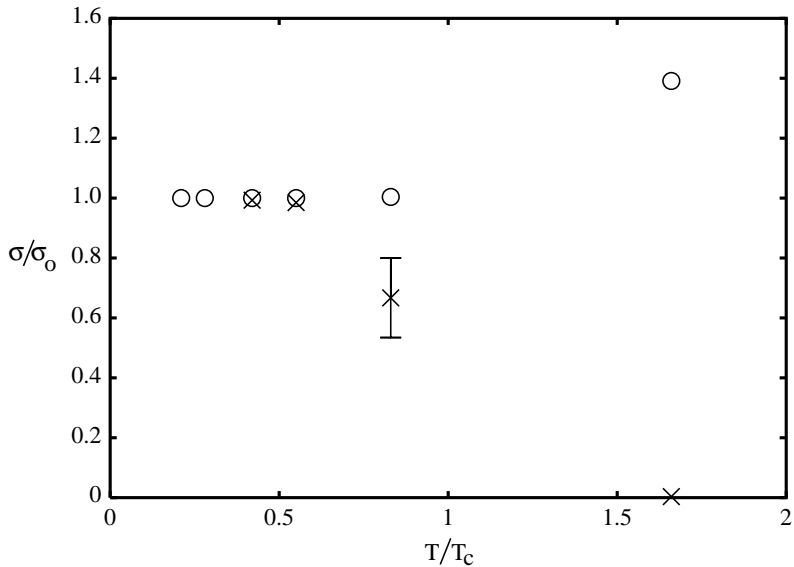}
\vspace{-2.5cm}
}
\caption{String tension between static color sources (crosses) and
spatial string tension (circles) as a function of temperature.}
\label{stt}
\end{figure}

Fig. \ref{stt} displays the result of string tension measurements on
$16^3 \times N_t $ lattices, as a function of temperature, for the choice 
of curvature coefficient $c=0.24$. Whereas the quantitative behavior of 
the static quark string tension, represented by the crosses in 
Fig. \ref{stt}, has been fitted using the freedom in the choice
of $c$, the so-called spatial string tension $\sigma_{s} $, represented 
by the circles in Fig. \ref{stt}, can now be predicted. In the
high-temperature regime, it begins to rise with temperature; the
value obtained at $T=1.67 \, T_C $, namely 
$\sigma_{s} (T=1.67 \, T_C ) = 1.39 \sigma_{0} $, corresponds to
within 1\% with the value measured in full $SU(2)$ Yang-Mills 
theory.\cite{karsch}

\begin{figure}[h]
\centerline{
\epsfxsize=5cm
\epsffile{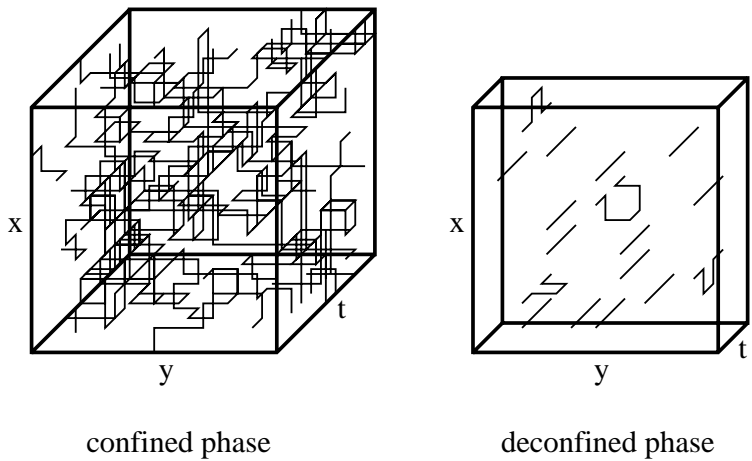}
\vspace{0.4cm}
}
\caption{Typical vortex configurations in the confined and the deconfined
phases.}
\label{pha}
\end{figure}

The confined and deconfined phases can alternatively be characterized by
the percolation properties \cite{selprep} of the vortex clusters in space 
slices of the lattice universe, i.e. slices defined by keeping one space 
coordinate fixed. In such a slice of space-time, vortices are represented 
by closed lines, cf. Fig. \ref{pha}. In the confined phase, these lines
percolate throughout (sliced) space-time, whereas in the deconfined phase,
they form small, isolated clusters, which, more specifically, wind around
the universe in the Euclidean time direction (and are closed by virtue of
the periodic boundary conditions). Also, simple heuristic arguments can be
given \cite{selprep} which explain why confinement should be associated 
with vortex percolation. The percolation characteristics of the surfaces 
in the vortex model precisely mirror the ones found for vortex structures 
extracted from full lattice Yang-Mills configurations.\cite{tlang}

\section{Topology}
\label{topsec}
Next to the confinement phenomenon, the topological properties of the 
Yang-Mills ensemble constitute an important nonperturbative aspect of
strong interaction theory, as already highlighted in section \ref{intro}. 
These properties are encoded in the topological winding number
\begin{equation}
Q=\frac{1}{32\pi^{2} } \int d^4 x \, \epsilon_{\mu \nu \lambda \tau } \
\mbox{Tr} \ F_{\mu \nu } F_{\lambda \tau } \ .
\end{equation}
In view of this expression, to generate a nonvanishing topological 
density at a certain point in space-time, the field strength there
must have nonvanishing tensor components such that the corresponding
Lorentz indices span all four space-time directions. For example,
a nonvanishing $F_{12} $ in conjunction with a nonvanishing $F_{34} $
would fulfil this requirement. In order to generate nonvanishing
$F_{12} $, a vortex surface segment (locally) running in 3-4 direction 
is necessary, as discussed in section \ref{defsec}; conversely, to
generate nonvanishing $F_{34} $, one needs a surface segment running
in 1-2 direction. Thus, a nontrivial contribution to the topological 
winding number is e.g. generated by a self-intersection point of the 
vortex surfaces, where a segment running in 3-4 direction intersects
a segment running in 1-2 direction. Quantitatively,\cite{cont} the 
contribution specifically of such a self-intersection point has modulus 
$1/2$. In general, all singular points
of a surface configuration contribute, where a singular point is
defined as a point at which the set of tangent vectors to the surface
configuration spans all four space-time directions. Self-intersection
points are but the simplest example of such singular points; there
also exist writhing points, at which the surface is twisted in such a
way as to generate a singular point in the above sense. These writhing
points actually turn out to be statistically far more important than the
self-intersection points in the random surface ensemble discussed 
here.\cite{preptop}

As already indicated in section \ref{defsec}, the orientation of the
vortex surfaces enters the topological winding number, since it is encoded
in the sign of the field strength. This should be contrasted with
the Wilson loop, which is insensitive to the orientation.
The random surfaces of the vortex ensemble should not be considered to
be globally oriented (in fact, most configurations are not even globally
orientable); they in general should be composed of patches of alternating 
orientation. Given a vortex surface from the ensemble
defined by (\ref{actio}), it is straightforward to randomly assign
orientations to the plaquettes making up the surface. By biasing this
procedure with respect to the relative orientation of neighboring
plaquettes, different mean sizes of the oriented patches making up
the surface can be generated; equivalently, the density of patch
boundary lines can be adjusted. This density strictly speaking
constitutes an additional parameter of the model, which cannot be fixed
using the confinement properties due to the fact that the Wilson loop is
insensitive to the vortex orientation. A priori, one might expect e.g. the
topological susceptibility, the measurement of which is presented below, 
to depend on this parameter. This would mean that the topological 
susceptibility can possibly be fitted, but not predicted. 
In actual fact, it turns out that this quantity is
independent of the aforementioned density within the error bars. The 
reasons for this can be understood in detail in terms of the properties 
of the random surface ensemble.\cite{preptop} The measurement of the
topological susceptibility exhibited below therefore does represent a 
genuine quantitative prediction of the vortex model.

At first sight, given the surfaces, including their orientation, it may
seem straightforward to locate their singular points and thus 
evaluate the topological winding number. However, surfaces constructed
on a hypercubic lattice, as done in the present formulation of the
vortex model, contain ambiguities which would not appear if one were
dealing with arbitrary surfaces in continuous space-time. For one,
the lattice surfaces in general self-intersect not at points, but along
whole intersection lines; in an ensemble of arbitrary continuum surfaces,
such configurations would represent a set of measure zero. Likewise, the
boundaries of the oriented patches making up the surfaces discussed
above coincide with singular points of the surfaces with a finite
probability; again, such configurations would represent a set of
measure zero in an ensemble of arbitrary continuum surfaces.
These ambiguities are resolved by interpreting the surfaces on the
hypercubic lattice as coarse-grained representations of other surfaces 
which slightly deviate from the former ones. Physically, this is
consistent; in view of the thickness of the vortices, the location of 
the associated surfaces is only defined up to arbitrary deviations
shorter than this thickness. Thus, in practice, the initial vortex
surfaces can be transferred onto a finer lattice, and arbitrary
short-range deformations can be carried out on the surfaces such as
to eliminate all the ambiguities highlighted above. Then the topological
winding number can indeed be extracted unambiguously.

Note therefore that the calculation of the topological properties
within the vortex model on a hypercubic lattice is afflicted with
difficulties which are reminiscent of the ones occuring in standard
lattice gauge theory. The elimination of ambiguities discussed above
is analogous to an ``inverse blocking'' prescription for Yang-Mills
link configurations.

\begin{figure}[ht]
\centerline{
\epsfxsize=11cm
\epsffile{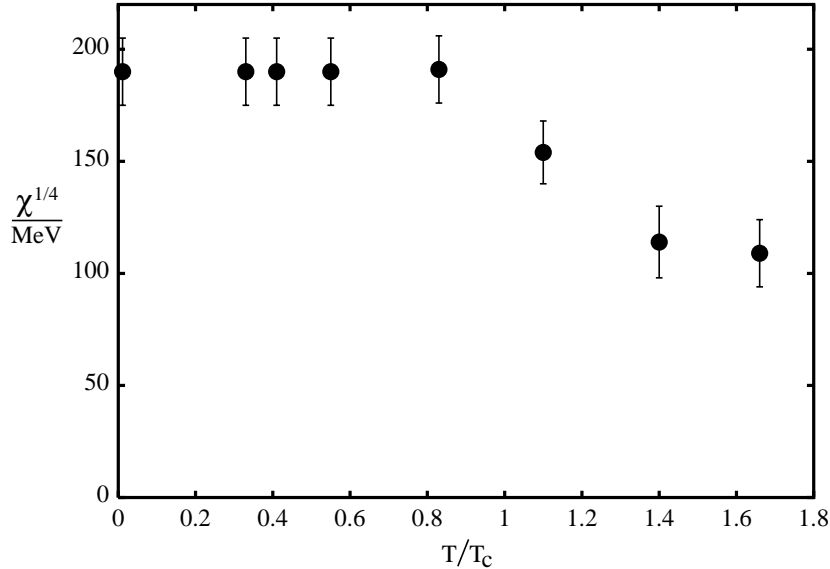}
}
\caption{Fourth root of the topological susceptibility as a function of
temperature.}
\label{chifig}
\end{figure}

Having thus obtained an algorithm allowing to evaluate the topological
winding number $Q$ of vortex surface configurations on a hypercubic lattice,
it is possible to measure the topological susceptibility 
$\chi = \langle Q^2 \rangle /V$ of the vortex ensemble, where $V$
denotes the space-time volume under consideration. The result is
exhibited in Fig. \ref{chifig} as a function of temperature. 

Quantitatively, this result is compatible with measurements in full
Yang-Mills theory.\cite{digia} The vortex model thus provides, within
one common framework, a simultaneous, consistent description of both the
confinement properties as well as the topological properties of the
$SU(2)$ Yang-Mills ensemble. It should remarked that, also as far as the 
vortex structures extracted from lattice Yang-Mills configurations are
concerned, evidence has been obtained that these structures encode
the topological characteristics of the gauge fields: Removing all
vortices from a lattice gauge configuration has the effect of transferring 
the configuration to the topologically trivial ($Q=0$) sector.\cite{forc1}

\section{Outlook}
\label{outsec}
One obvious generalization of the present work is the treatment of
$SU(3)$ color. The vortex model appropriate for this color group
will in some respects exhibit qualitatively different behavior. 
Since there are two nontrivial center elements in the $SU(3)$ group, 
namely the phases $e^{\pm i 2\pi /3} $, one will have to distinguish 
two distinct vortex fluxes. The main qualitative difference in the 
topology of vortex configurations will be the presence of vortex 
branchings. In the $SU(3)$ theory, there are branchings which respect 
flux conservation; a vortex carrying one type of vortex flux can split 
into two vortices carrying the other type of vortex flux. This difference 
in the topological character of the configurations is expected to lead 
to a change in the order of the deconfinement phase transition from 
second order for $SU(2)$ to first order for $SU(3)$, as observed in full 
Yang-Mills lattice experiments.\cite{kopo}

To render the vortex model a useful quantitative tool, it is furthermore
important to couple the vortices to quark degrees of freedom. Foremost,
it must be ascertained whether the vortex ensemble induces the
spontaneous breaking of chiral symmetry to a sufficient degree; i.e.,
the chiral condensate must be evaluated quantitatively.
Technically, this implies constructing the Dirac operator in a vortex
background. In this respect, the vortex picture has an important
advantage to offer. As hinted in section \ref{defsec}, any arbitrary 
vortex surface can be associated with a continuum gauge field; this
trivially includes surfaces which happen to be made up of elementary
squares, e.g. plaquettes on a hypercubic lattice. Therefore, the
Dirac operator can be constructed directly in the continuum, and some
of the difficulties associated with lattice Dirac operators, such as
fermion species doubling, may be avoidable. From a physical point of
view, it is tempting to speculate that a model with the correct
topological properties should also correctly account for the chiral 
condensate, in view of the corresponding success of instanton 
models \cite{inst} in describing the spontaneous breaking of chiral 
symmetry.

\section*{Acknowledgments} 
M.E.~and H.R.~gratefully acknowledge financial support by Deutsche 
Forschungsgemeinschaft under grants DFG En 415/1-1 and DFG Re 856/4-1,
respectively. M.F.~is supported by Fonds zur F\"orderung der
Wissenschaftlichen Forschung under P11387-PHY.

\end{document}